\documentclass[letterpaper, preprint, paper,11pt]{IAA-AAS}	% for preprint proceedings

\usepackage{bm}
\usepackage{amsmath,amsthm,amssymb}
\usepackage{epsfig} % for postscript graphics fileshttps://www.overleaf.com/project/5ea73852bcebd10001c43669
\usepackage{epstopdf}
\usepackage{color}
\usepackage{algpseudocode,algorithm,algorithmicx}
\usepackage{subfigure}
\usepackage[colorlinks=true, pdfstartview=FitV, linkcolor=black, citecolor= black, urlcolor= black]{hyperref}
\usepackage{overcite}
\usepackage{footnpag}			      	% make footnote symbols restart on each page
\usepackage[dvipsnames]{xcolor}
\usepackage[deletedmarkup=sout,authormarkup=superscript]{changes}

\definechangesauthor[name={AJ}, color=NavyBlue]{AJ}
\definechangesauthor[name={Dominic}, color=RedViolet]{DLM}

\newcommand\blfootnote[1]{%
  \begingroup
  \renewcommand\thefootnote{}\footnote{#1}%
  \addtocounter{footnote}{-1}%
  \endgroup
}

\PaperNumber{-XX-XX}

\newtheorem{rmk}{Remark}

\begin{document}

\title{Suboptimal Nonlinear Model Predictive Control Strategies for Tracking Near Rectilinear Halo Orbits}

\author{Andrew W. Berning Jr.\thanks{PhD Candidate, Aerospace Engineering, University of Michigan, 1221 Beal Avenue, Ann Arbor, MI 48109.},  
Dominic Liao-McPherson\thanks{Research Fellow, Aerospace Engineering, University of Michigan, 1221 Beal Avenue, Ann Arbor, MI 48109.},
Anouck Girard\thanks{Associate Professor, Aerospace Engineering, University of Michigan, 1221 Beal Avenue, Ann Arbor, MI 48109.}, 
 and Ilya Kolmanovsky\thanks{Professor, Aerospace Engineering, University of Michigan, 1221 Beal Avenue, Ann Arbor, MI 48109.}\blfootnote{This research is supported by the National Science Foundation Award Number CMMI 1562209.}
}

\maketitle{}

\begin{abstract}
Near Rectilinear Halo Orbits (NRHOs), a subclass of halo orbits around the L1 and L2 Lagrange points, are promising candidates for future lunar gateways in cis-lunar space and as staging orbits for lunar missions.  
Closed-loop control is beneficial 
to compensate orbital perturbations and potential instabilities while maintaining spacecraft on NRHOs and performing relative motion maneuvers.
This paper investigates the use of nonlinear model predictive control (NMPC) coupled with low-thrust actuators for station-keeping on NRHOs. It is demonstrated through numerical simulations that NMPC is able to stabilize a spacecraft to a reference orbit and handle control constraints. Further, it is shown that the computational burden of NMPC can be managed using specialized optimization routines and suboptimal approaches without jeopardizing closed-loop performance.
\end{abstract}

\section{Introduction}

It has been more than 40 years since a human last traveled beyond low earth orbit (LEO) on the Apollo 17 spacecraft in 1972. Recently, there has been renewed interest in human exploration of the solar system. In particular, cis-lunar space has emerged as a focus area as evidenced by the Global Exploration Roadmap \cite{hufenbach2015international,laurini2015global} and the NASA Artemis program\cite{woodard2009artemis}, including the 2024 lunar landing goal\cite{nasa_lunar_landing}. Operations in cis-lunar space will support space-based facilities for robotic and human missions to the Moon and eventually to destinations such as asteroids or Mars.

Near Rectilinear Halo Orbits (NRHOs) have emerged as promising candidates for a long term lunar gateway module and/or as staging orbits between LEO and low lunar orbit\cite{zimovan2017near,whitley2016options}. NRHOs are limit cycles near the co-linear L1 and L2 Lagrange points. Certain NHROs possess several useful properties including the existence of low-energy transfer orbits \cite{bury2020landing}, good stability characteristics, unobstructed views of Earth, and favorable resonance properties that allow them to avoid eclipses \cite{guzzetti2017stationkeeping}. They were first identified in the Circular Restricted Three Body Problem (CR3BP) and are periodic natural motion trajectories in the simplified setting of the CR3BP\cite{howell1984almost}.

While certain NRHOs are stable or nearly stable in restricted three body systems\cite{zimovan2017near}, in reality maintaining the spacecraft on them is challenging due to perturbations caused by gravitational forces from other celestial bodies, navigational errors, hardware limitations, solar radiation and magnetic forces. This provides the motivation for the development of active station-keeping algorithms that can stabilize the orbit despite these disturbances. Ideally, these algorithms should optimally balance tracking error with propellant/energy consumption. Computing actions/policies offline and uploading them reduces onboard computing requirements but may reduce robustness to disturbances. Alternatively, computing actions online can enhance the ability to react to changes and improve robustness. In particular, Model Predictive Control (MPC) is a promising methodology for online optimal control that can systematically account for constraints, nonlinearities, and both trajectory tracking and fuel minimization requirements\cite{rawlings2009model,grune2017nonlinear,ellis2014tutorial}. However, online optimization approaches can be expensive from a computational or power consumption perspective.

There is a growing body of literature on station-keeping for NRHOs. Dynamical  systems theory based techniques using Cauchy-Green Tensors and X-Axis Crossing methods have, in particular, been investigated \cite{guzzetti2017stationkeeping,davis2017orbit}.
Set-invariance and analytical minimum energy-based solutions to a linearized problem that account for measurement uncertainty have been proposed\cite{muralidharan2020control}, and linear MPC and linear quadratic regulator (LQR) approaches have been developed\cite{kalabic2015station}. There is also interest in station-keeping for other halo orbits about the L1 and L2 points. The Optimal Continuation Strategies method, based on 2-point boundary value problems, was validated in-flight during NASA's ARTEMIS mission\cite{folta2014earth} which flew longer period halo orbits around the L1 and L2 points. Discrete time sliding mode control has been applied to station-keeping on Halo and Lissajous orbits around L2 \cite{lian2014station}. Analytical methods for station-keeping on Halo orbits in the CR3BP using the continuous-time linear quadratic regulator\cite{breakwell1974station} (LQR), Floquet theory\cite{simo1987optimal}, and invariant manifolds\cite{gomez1986station} have also been investigated.

In this paper, we investigate the use of nonlinear model predictive control (NMPC) for station keeping on NRHOs using a low-thrust propulsion system. We show that NMPC is able to successfully maintain spacecraft flight along a  halo orbit, satisfy thrust constraints, and demonstrates a high degree of robustness to disturbances and model mismatch. Moreover, while low-thrust actuators are well suited for long duration station-keeping, they require more frequent control updates, leading to more demanding onboard computing requirements. As such, we leverage new optimization algorithms\cite{liaomcpherson2020fbstab} and suboptimal MPC\cite{liaomcpherson2020tdo}, to demonstrate that it is possible to meet closed-loop performance requirements using little computational power. Our approach contrasts with existing literature which often assumes impulsive thrusters and uses control methodologies based on linearized models. By demonstrating the computational feasibility of NMPC for this problem, we open the door to future research exploiting the flexibility, i.e., nonlinear cost functions and constraints, of NMPC to optimize high-level objectives such as minimizing propellant consumption.

\section{System Modelling}
When considering spaceflight in the cis-lunar flight regime, it is most natural to consider the restricted three body problem in which the third body is of negligible mass compared to the two primary bodies (as is the case for a spacecraft, the moon, and the earth)\cite{broucke1969stability}. The two restricted three body problems considered in this work are the circular restricted three body problem (for reference trajectory generation and control prediction model), in which the primary and secondary bodies are assumed to travel in circular orbits about their barycenter, and the elliptical restricted three body problem (for simulation and validation), in which the orbits of the primary and secondary bodies are assumed to have nonzero eccentricities. 

For both systems of equations, described below, the primary and secondary bodies lie along the $x$ axis with the barycenter at the origin. The $z$ axis points in the direction of angular momentum of the primary-secondary system, and the $y$ axis completes the orthogonal frame. 

\subsection{The Elliptical Restricted Three Body Problem (ER3BP)}
\begin{figure}[htbp!]
	\centering
	\includegraphics[width=4in]{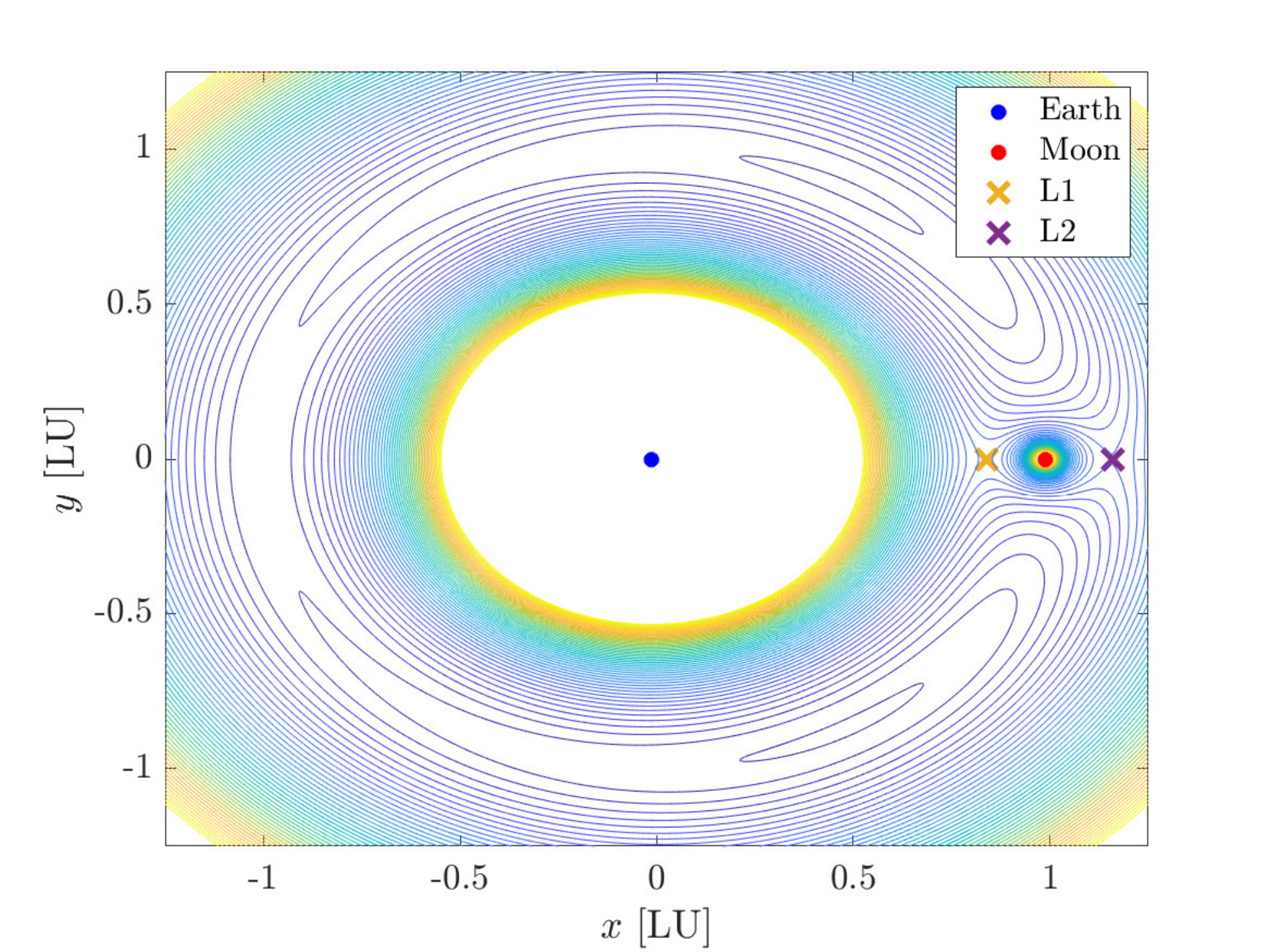}
	\caption{Contour plot of the pseudo-potential \eqref{eq:pseudopotential} with Lagrangian points L1 and L2 marked.}
	\label{fig:pseudo}
\end{figure}

The equations of motion of the spacecraft in the pulsating frame of ER3BP and in non-dimensionalized distance and time units are given by:
\begin{subequations} \label{eq:er3bp}
\begin{align}
    x'' - 2y' &= \frac{1}{1+e \cos(\theta)} \frac{\partial U}{\partial x} + u_x\\
    y'' + 2x' &=  \frac{1}{1+e \cos(\theta)}  \frac{\partial U}{\partial y} + u_y \\
    z''+ z &= \frac{1}{1+e \cos(\theta)} \frac{\partial U}{\partial z} + u_z
\end{align}
\end{subequations}
where $(\cdot)'$ denotes differentiation with respect to the true anomaly of the primaries $\theta$, $u = (u_x,u_y,u_z)$ are the control accelerations provided by thrusters and
\begin{equation} \label{eq:pseudopotential}
   U(x,y,z) = \frac12 (x^2 + y^2) + \frac{1-\mu}{\|(x+\mu,y,z)\|_2} + \frac{\mu}{\|(x-1+\mu,y,z)\|_2},
 \end{equation}
 is the pseudo-potential function, shown in Figure \ref{fig:pseudo}. Note that the independent variable in \eqref{eq:er3bp} is the true anomaly which is related to time, $t$ by 
\begin{equation}
  t' = \frac{1}{(1+e \cos(\theta))^2}. \\ % I changed t^' to t, is this correct (IK)
\end{equation}

In the absence of eccentricity, i.e., with $e = 0$, the ER3BP reduces to the circular restricted three body problem (CR3BP):
\begin{subequations}
\begin{align}
x'' - 2y' = &~ \frac{\partial U}{\partial x} + u_x\\ 
y'' + 2x' =&  \frac{\partial U}{\partial y} + u_y,\\
z'' + z =&  \frac{\partial U}{\partial z}  + u_z.
\end{align}
\end{subequations}
Note that in the circular case, the rate of change of time with respect to true anomaly is unity and so, barring wrap-around issues ($\theta \in [0,2 \pi]$) , $t = \theta$. 

The equations of motion can be simplified by introducing the velocity $v = (x',y',z')$ and position $r = (x,y,z)$ vectors. Then \eqref{eq:er3bp} becomes
\begin{subequations}  \label{eq:eom_compact1}
\begin{equation}
  \begin{bmatrix}
    r'\\v'
  \end{bmatrix} = \begin{bmatrix}
    0 & I \\
    A_{21} & A_{22}
  \end{bmatrix} \begin{bmatrix}
    r\\v
  \end{bmatrix} + \begin{bmatrix}
    0\\ I
  \end{bmatrix} \left(\frac{\nabla U(r) }{1+e \cos(\theta)}+ u\right)
\end{equation}
where
\begin{equation}
  A_{21} = \begin{bmatrix}
    0 & 0 & 0\\
    0 & 0 & 0\\
    0 & 0 & 1
  \end{bmatrix}, \quad A_{22} = \begin{bmatrix}
    0 & 2 & 0\\
    -2 & 0 & 0\\
    0 & 0 & 0
  \end{bmatrix}.
\end{equation}
\end{subequations}
Finally, introducing the state vector $\xi = (r,v)$, \eqref{eq:eom_compact1} can be written compactly as
\begin{equation} \label{eq:continuous_time_model}
  \xi' = f_c(\theta,\xi,u,e).
\end{equation}

\subsection{Halo Orbits}

Halo orbits are families of periodic orbits near the collinear Lagrange points in the CR3BP\cite{howell1984almost}. Members of these orbital families that occur closer to the secondary body begin to exhibit properties of NRHOs, traveling very nearly in a plane normal to the orbital plane of the primaries. 

\begin{figure}[htbp!]
	\centering
	\includegraphics[width=4in]{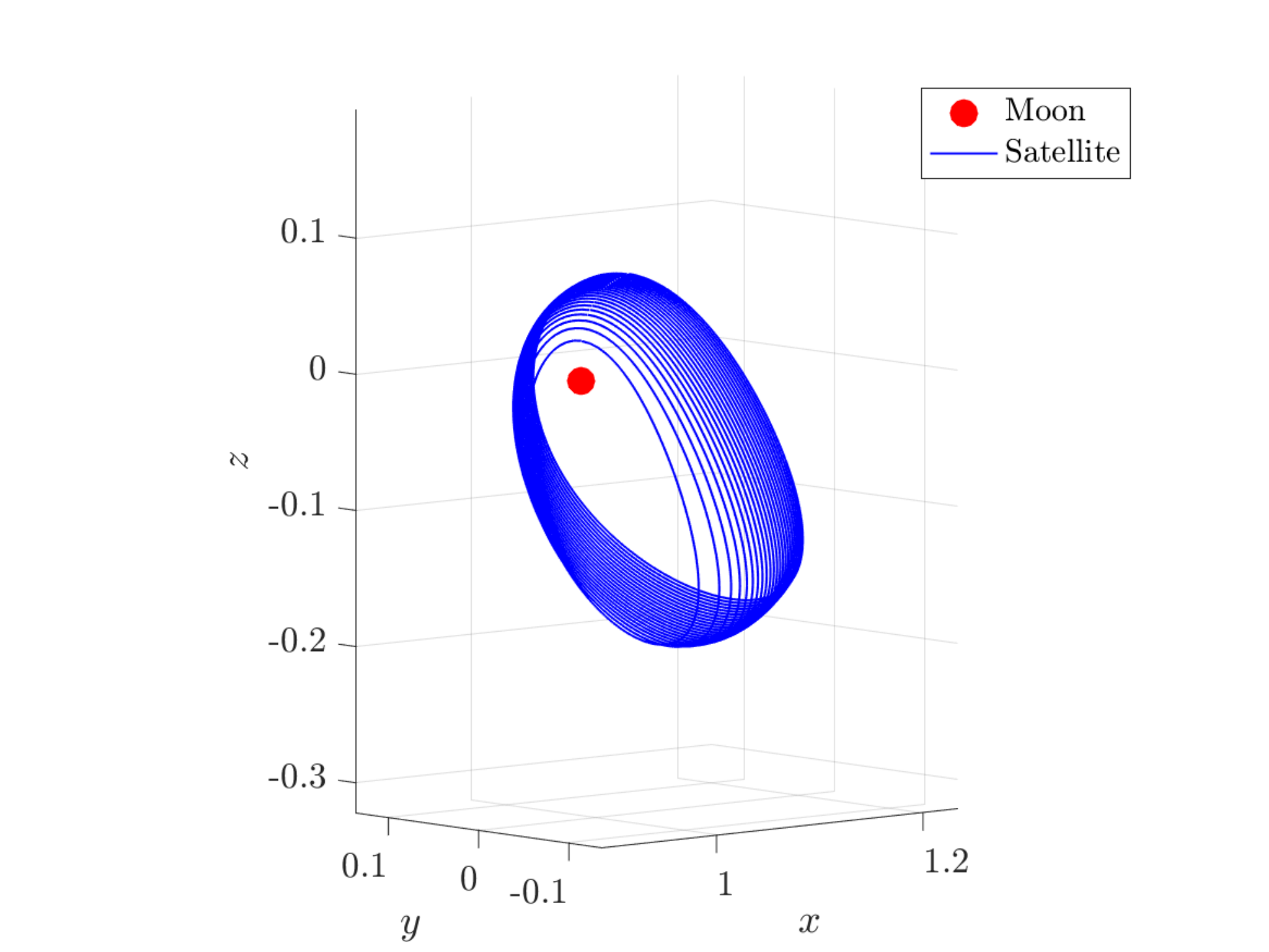}
	\caption{Family of Earth-Moon L2 halo orbits in CR3BP.}
	\label{fig:halo_sweep}
\end{figure}

When considering the construction of these orbits in CR3BP we exploit the following property: the velocities $x'$ and $z'$ are equal to zero when the orbit crosses the $x \mbox{-} z$ plane ($y=0$) at apoapsis and periapsis. Additionally, orbits are symmetric across the $x \mbox{-} z$ plane, so if we assume initial conditions of $X_0 = [x_0, 0, z_0, 0, y'_0, 0]^{\rm T}$ and integrate until the trajectory again crosses the $x \mbox{-} z$ plane at $X_f = [x_f, 0, z_f, x'_f, y'_f, z'_f]^{\rm T}$, we only need to enforce $ x'_f=z'_f=0$ to ensure a periodic orbit. In this work, a single-shooting method is used to solve this problem, iterating on the initial conditions $z_0$ and $y'_0$ for a given $x_0$ until the condition $ x'_f=z'_f=0$ is met to within some specified tolerance. Finally, in an attempt to reduce the discontinuity when a controller is tracking the reference trajectory over multiple orbital periods, the state at each time instant is shifted linearly such that the shift at $X_0$ is zero and the shift at $X_f$ is such that $X_f = X_0$. This results in a reference trajectory that is no longer a natural motion trajectory, but it eliminates the aforementioned discontinuity. A family of periodic orbits about L2 is shown in Figure \ref{fig:halo_sweep}.

The single-shooting method utilized in this work is limited to finding periodic orbits in system \eqref{eq:er3bp} only for $e=0$. Future work will include the use of a more sophisticated multiple-shooting and
homotopy
methods to enable finding periodic reference trajectories in ER3BP for $e\neq0$. 

\section{Control Design}
This section describes the proposed controller. An MPC controller has three constituent components, a prediction model used to evaluate the impact of control actions, an optimal control problem (OCP) formulation encapsulating the control problem, and an implementation strategy for solving the OCPs online.
\subsection{Prediction Model}
MPC uses a prediction model to estimate the response of the system to control actions. Such a model is typically of the form
\begin{equation}
  \xi_{k+1} = f(\theta_k,\xi_k, u_k),
\end{equation}
where $\xi_k = \xi(\theta_k)$ and $u_k = u(\theta_k)$. We derive a suitable prediction model in this form by discretizing the CR3BP, i.e., \eqref{eq:eom_compact1} with $e = 0$. The control signal is held constant over each interval $[\theta_k,\theta_{k+1}]$ so the exact model is given by the state transition equation
\begin{equation} \label{eq:prediction_stm}
  \xi_{k+1} = \phi(\theta_k,\xi_k,u_k) = \xi_k + \int_{\theta_k}^{\theta_{k+1}} f_c(\theta,\xi(\theta) ,u_k,0) ~d\theta.
\end{equation}
We approximate \eqref{eq:prediction_stm} numerically using a standard 4$^{th}$ order Runge-Kutta (RK4) method:
\begin{gather} \label{eq:rk4}
\xi_{k+1} = f(\theta_k,\xi_k,u_k) = \xi_k + \frac{\Delta \theta}{\|b\|_1} \sum_{i=1}^4 b_i~k_i(\theta_k,\xi_k,u_k),\\ k_i= f_c(\theta_k+ \Delta \theta~ c_i, \xi_k + \Delta \theta~ a_i~k_{i-1}, u_k,0), ~k_0 = 0, \nonumber
\end{gather}
with a uniform step size $\Delta \theta > 0$ and where $a = (0,0.5,0.5,1)$, $b = (1,2,2,1)$, and $c = (0,0.5,0.5,1)$ are the coefficients of the method.

\begin{rmk}
Throughout this paper, the CR3BP is used as the prediction model.
\end{rmk}

\subsection{Optimal Control Problem Formulation} \label{sec:OCP_formulation}
In MPC the feedback law is defined implicitly through the solution of a receding horizon optimal control problem (OCP). Let $N > 0$ be the length of the prediction horizon, $i$ be the index along the prediction horizon, and $k$ be the discrete true anomaly index. We use the notation $\xi_{i|k}$ to denote the predicted state $i$ steps into the prediction horizon at the sampling instant $k$ and denote the planned control actions $u_{i|k}$ analogously. The OCP formulation is then
\begin{subequations} \label{eq:OCP}
\begin{alignat}{2}
	&\min_{\tilde{\xi}, \tilde{u}} ~~~ &&||\xi_{i|k} - \bar{\xi}_{i|k}||_Q^2 + \frac12\sum_{i=0}^{N-1} ||\xi_{i|k} - \bar{\xi}_{i|k}||_Q^2 + ||u_{i|k}||_R^2 \label{eq:OCP_cost} \\ 
	&s.t. ~~ &&\xi_{0|k} = \xi_k, \\
	& &&\xi_{i+1|k} = f(\theta_{i|k},\xi_{i|k},u_{i|k}), ~~i = 0,\dots,N{-}1, \\
	& &&\|u_{i|k}\|_\infty \leq u_{max}, ~~i = 0,\dots,N,
\end{alignat}
\end{subequations}
where $\theta_{i|k} = \theta_k + i\Delta \theta$, $\tilde{\xi} = (\xi_{0|k}, \ldots, \xi_{N|k})$ and $\tilde{u} = (u_{0|k},\ldots, u_{N-1|k})$ are the optimization variables, $\bar{\xi}_{i|k}$ is the desired state, $u_{max} > 0$ is an upper bound on the input acceleration, $f$ is defined in \eqref{eq:rk4}, and $Q = Q^T \succ 0$ and $R = R^T \succ 0$ are weighting matrices. The control input is then $u_k = u^*_{0|k}$ where $(\cdot)^*$ denotes a minimizer of \eqref{eq:OCP}.

\subsection{Controller Implementation}
\begin{figure}[htbp]
  \centering
  \includegraphics[width=0.6\textwidth]{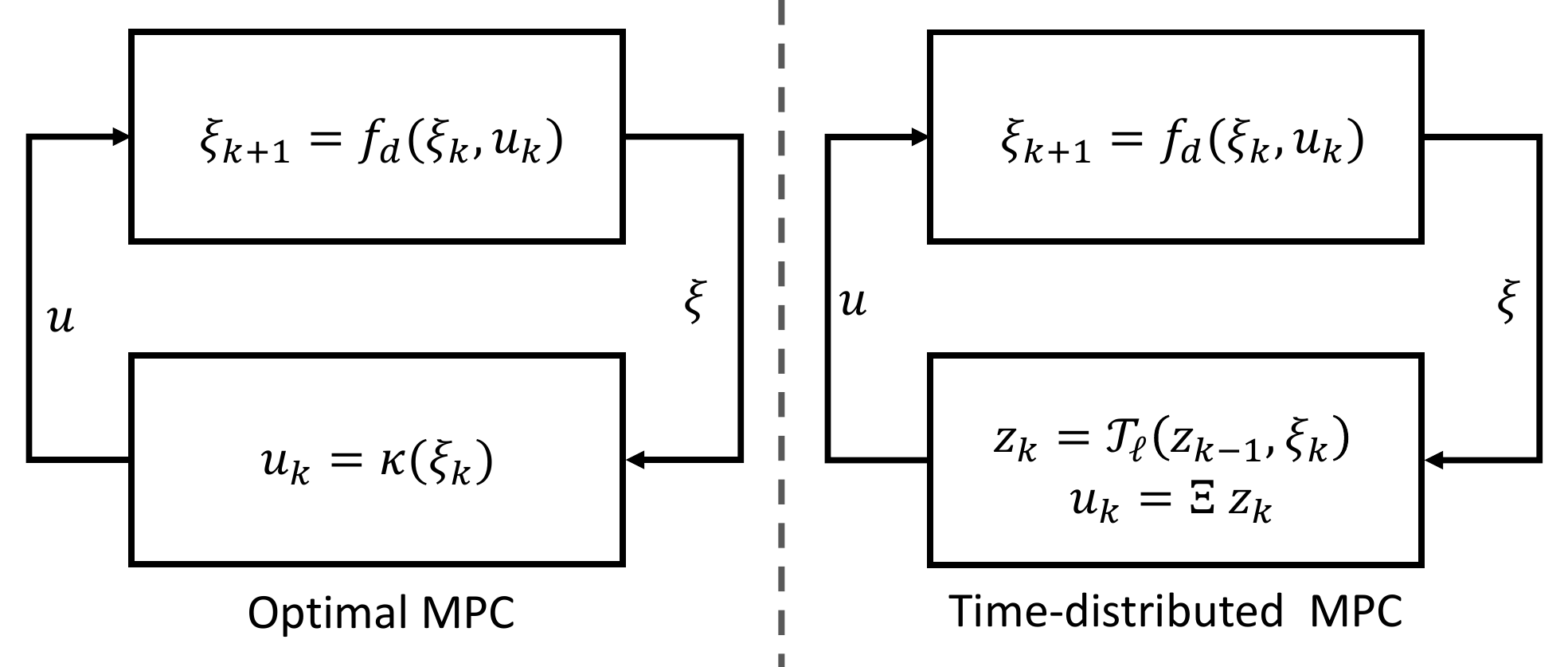}
  \caption{A comparison between optimal MPC, which is a static feedback law, and time-distributed MPC, which is a dynamic compensator. The operator $\mathcal{T}_M$ represents a fixed number of SQP iterations and $\Xi$ selects the control input from the full solution.}
  \label{fig:tdo}
\end{figure}

An efficient method for solving \eqref{eq:OCP} is essential for implementing NMPC. In this paper, we use a suboptimal variant of Sequential Quadratic Programming (SQP) algorithm that exploits the time sequential structure of MPC. The OCP \eqref{eq:OCP} can be written compactly as the nonlinear programming problem
\begin{subequations}
\begin{alignat}{2}
\underset{w}{\mathrm{min.}}& \quad &&\phi(w)\\
\mathrm{s.t.}&\quad &&g(w,\xi_k) = 0 \\
& &&h(w) \leq 0,
\end{alignat}
\end{subequations}
where $w = (\tilde{\xi},\tilde{u})$. The NLP is solved using the sequential quadratic programming iteration
\begin{equation}
  w_{i+1|k} = w_{i|k} + d^*, \quad \lambda_{i+1|k} = \lambda^*, \quad v_{i+1|k} = v^*,
\end{equation}
where $\lambda$ and $v$ are dual variables associated with the equality and inequality constraints, respectively, $z_{i|k} = (w_{i|k},\lambda_{i|k}, v_{i|k})$ is the solution estimate at the sampling instant $\theta_k$ after $i$ SQP iterations and $(d^*,\lambda^*,v^*)$ is the primal-dual solution to the following quadratic program (QP)
\begin{subequations} \label{eq:QP}
\begin{alignat}{2}
\underset{w}{\mathrm{min.}}& \quad &&d^T \nabla_w^2\phi(w_{i|k}) d + \nabla_w\phi(w_{i|k})^T d\\
\mathrm{s.t.}& && g(w_{i|k},\xi_k) + \nabla_w g(w_{i|k},\xi_k) d = 0,\\
& &&h(w_{i|k}) + \nabla_w h(w_{i|k}) d \leq 0.
\end{alignat}
\end{subequations}
In time-distributed optimization (TDO), the SQP algorithm is limited to $\ell > 0$ iterations and the final solution guess from previous timestep is used to warmstart the SQP algorithm, i.e., $z_{0|k+1} = z_{\ell|k}$. This leads to a coupled plant-optimizer system as shown in Figure~\ref{fig:tdo}. Under appropriate assumptions it is possible to show that TDO based MPC recovers the stability and robustness properties of optimal MPC using a finite number of iterations\cite{liaomcpherson2020tdo}.

Analytic derivatives, potentially coupled with symbolic optimization\cite{walker2016design}, are important for accelerating the SQP routine. The cost function and the inequality constraints are quadratic and linear functions so their derivatives are readily available. The derivative of the equality constraints, i.e., of the prediction model, are more complicated due to the RK4 integration scheme used in \eqref{eq:rk4}. Algorithm~\ref{algo:rk4_deriv} evaluates the sensitivities $\nabla_\xi f$ and $\nabla_u f$. Finally, the quadratic programming algorithm used to solve \eqref{eq:QP} significantly influences the overall computation time. We use the FBstab method\cite{liaomcpherson2020fbstab} which can exploit the sparsity structure of \eqref{eq:QP} and be easily warmstarted between SQP iterations.
\begin{algorithm}[h]
\caption{Sensitivities of RK4 Integration}
\label{algo:rk4_deriv}
\begin{algorithmic}[1]
\renewcommand{\algorithmicrequire}{\textbf{Inputs:}} 
\renewcommand{\algorithmicensure}{\textbf{Outputs:}} 
\Require $\Delta \theta$, $\xi$, $u$,$\theta$
\Ensure $f(\theta,\xi,u)$, $\nabla_\xi f(\theta, \xi,u)$, $\nabla_u f(\theta, \xi,u)$
\State $k_0 = 0$, $A_0 = 0$, $B_0 = 0$
\For{$i = 1,\ldots 4$}
\State $\theta_i = \theta + c_i \Delta \theta$
\State $\xi_i = \xi + a_i \Delta \theta k_{i-1}$
\State $k_i = f_c(\theta_i, \xi_i, u)$
\State $A_i = \nabla_\xi f_c(\theta_i,\xi_i,u) \left[I + a_i \Delta \theta A_{i-1}\right]$
\State $B_i = \nabla_\xi f_c(\theta_i,\xi_i,u) \left[a_i \Delta \theta B_{i-1}\right] + \nabla_u f_c(\theta_i,\xi_i,u)$
\EndFor
\State $f(\theta,\xi,u)= \xi + \frac{\Delta \theta}{\|b\|_1} \sum_{i=0}^4 b_i k_i$
\State $\nabla_\xi f(\theta, \xi,u) = I + \frac{\Delta \theta}{\|b\|_1} \sum_{i=0}^4 b_i A_i$
\State $\nabla_u f(\theta, \xi,u) = \frac{\Delta \theta}{\|b\|_1} \sum_{i=0}^4 b_i B_i$
\end{algorithmic}
\end{algorithm}
The plant and controller are both implemented in Simulink 2019b MATLAB function blocks and compiled into \texttt{C} code. We use a MATLAB implementation of FBstab\footnote{\url{https://github.com/dliaomcp/fbstab-matlab}} that is compatible with automatic code generation.

\section{Simulation Results}

The specific parameters used for simulations in this work are presented in Table \ref{tbl:params}. Simulations are performed under white Gaussian process noise with zero mean and variance $\tau$. 
%The spacecraft parameters and halo orbit \dlm{AJ is this true?} are roughly representative of current plans for the lunar gateway system. 
The thrust limits and mass used in simulation are in line with expected capabilities of the Advanced Electric Propulsion System \cite{herman2018overview}.

\begin{table}[h!]
  \centering
  \begin{tabular}{||c c c||} 
    \hline
    $m$ & $10,000~kg$ & Spacecraft mass \\ 
$ N$ & $35 $ & OCP horizon length \\ 
$M$ & $3$ & Maximum number of SQP iterations \\
$ \Delta \theta $ & $ 0.01~rad$ & Discrete step size \\ 
$  \mu$ & $  0.012$ & Mass ratio \\ 
$ e $ & $  0.055 $ & Eccentricity \\ 
$  \tau $ & $ 10^{-10}~\frac{km}{s^2} $ & Process noise variance \\ 
$ Q $ & $10^3 \begin{bmatrix}
10 \mathbb{I}_3 & 0_3 \\
0_3 & \mathbb{I}_3
\end{bmatrix}$ & Cost function weighting \\
$R$ & $\mathbb{I}_3$  & Cost function weighting \\
$u_{max}$ & $2000~mN$ & Max control constraint \\
    \hline
  \end{tabular}
  \caption{Simulation parameters}
  \label{tbl:params}
\end{table}

\begin{figure}[htbp!]
  \centering
  \includegraphics[width=\textwidth]{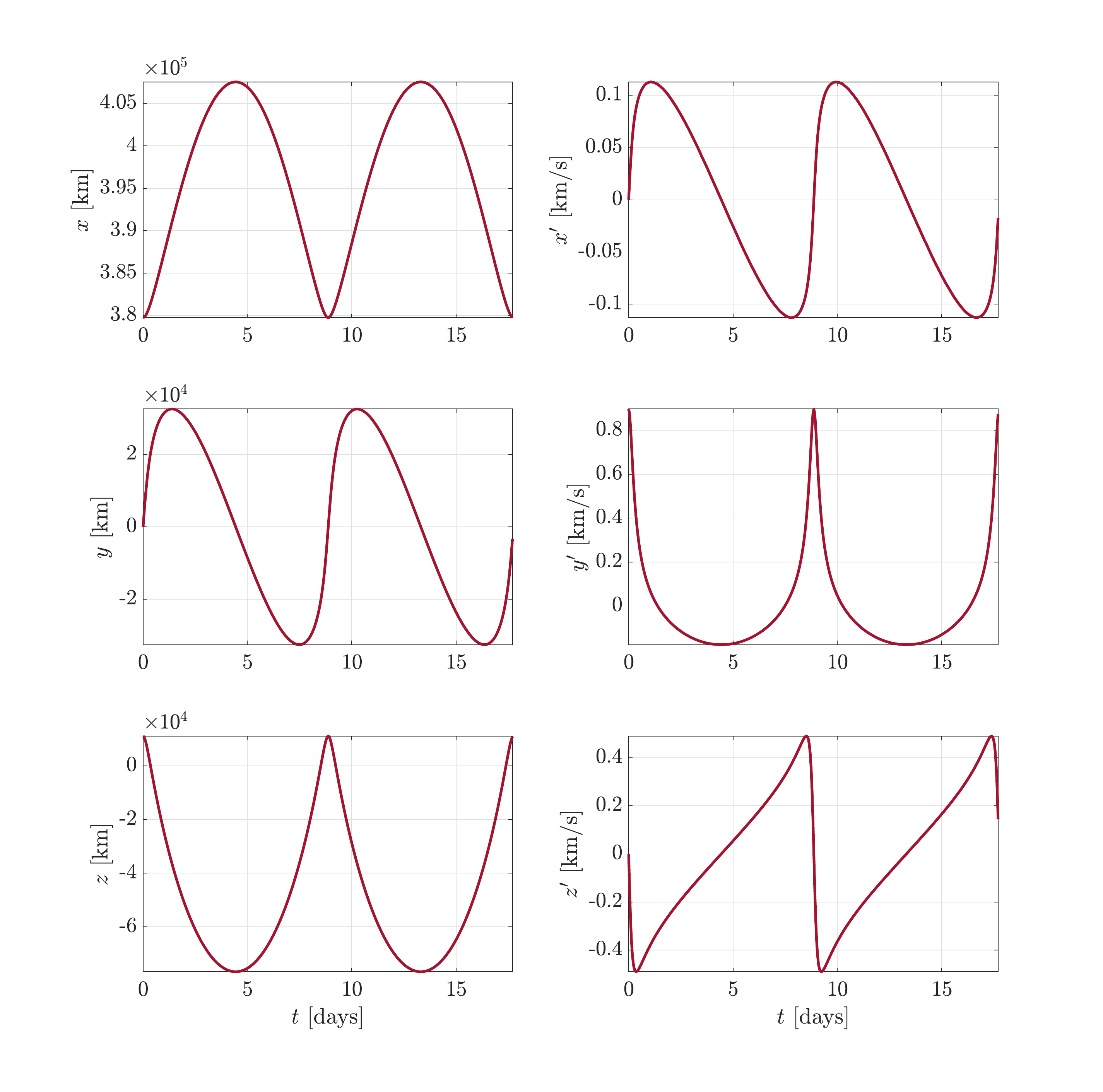}
  \caption{State trajectories for spacecraft in CR3BP tracking halo orbit.}
  \label{fig:states_cval}
\end{figure}

\begin{figure}[htbp!]
  \centering
  \includegraphics[width=\textwidth]{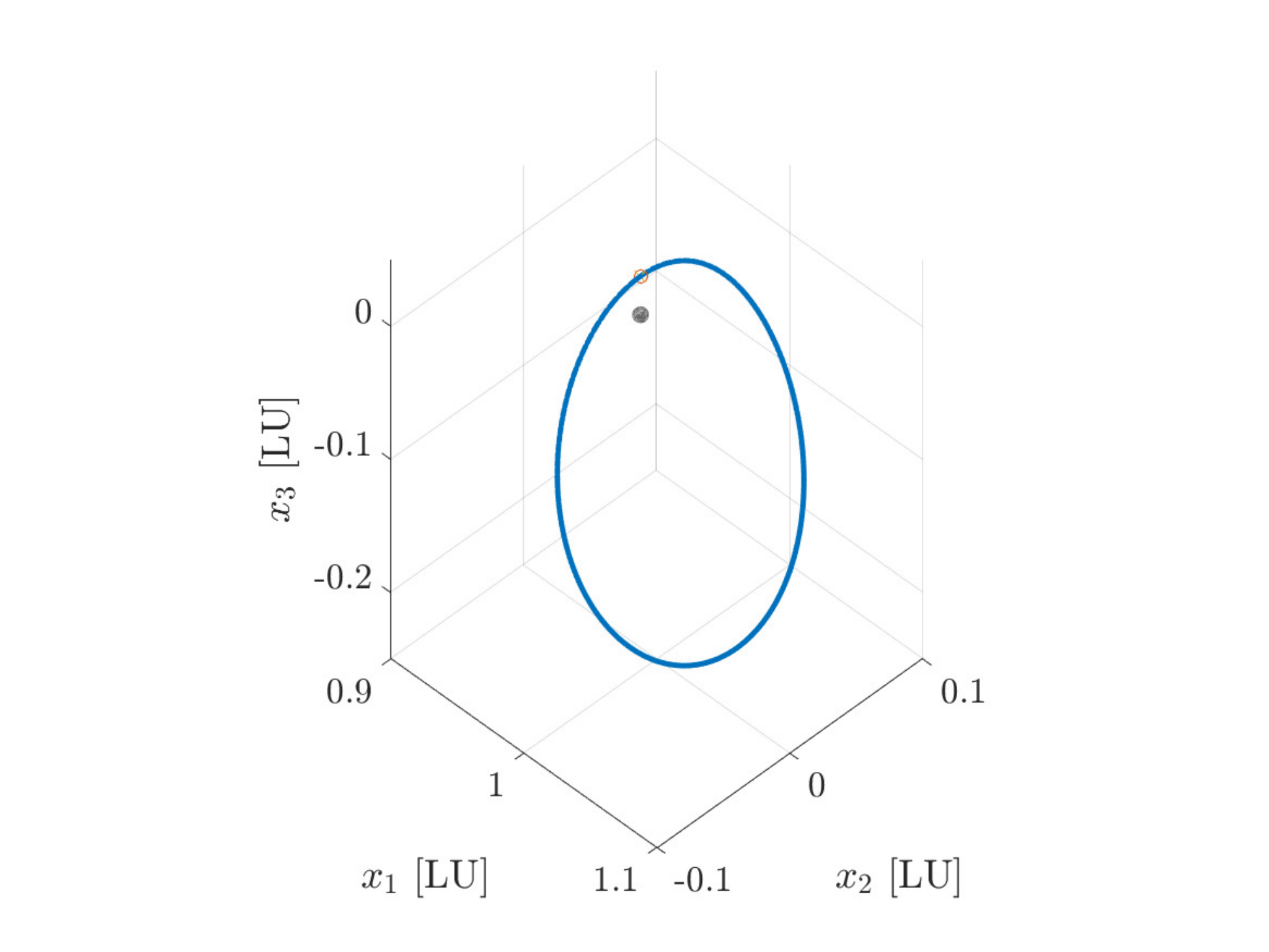}
  \caption{Spacecraft trajectory in CR3BP tracking halo orbit, displayed in non-dimensional length units [LU].}
  \label{fig:3dtraj_cval}
\end{figure}

\begin{figure}[htbp!]
  \centering
  \includegraphics[width=\textwidth]{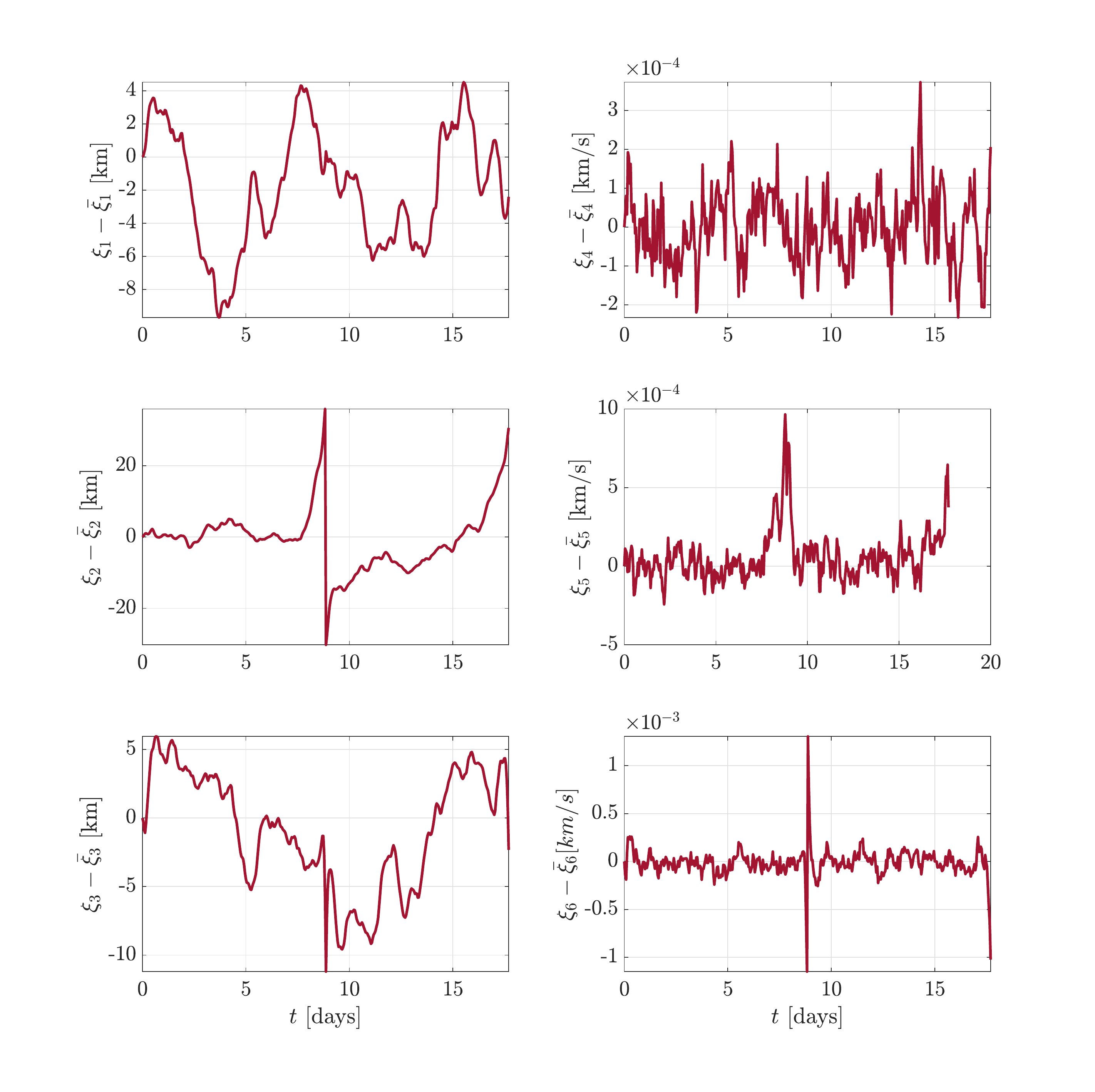}
  \caption{State error trajectories for spacecraft in CR3BP tracking halo orbit.}
  \label{fig:errors_cval}
\end{figure}

\begin{figure}[htbp!]
  \centering
  \includegraphics[width=\textwidth]{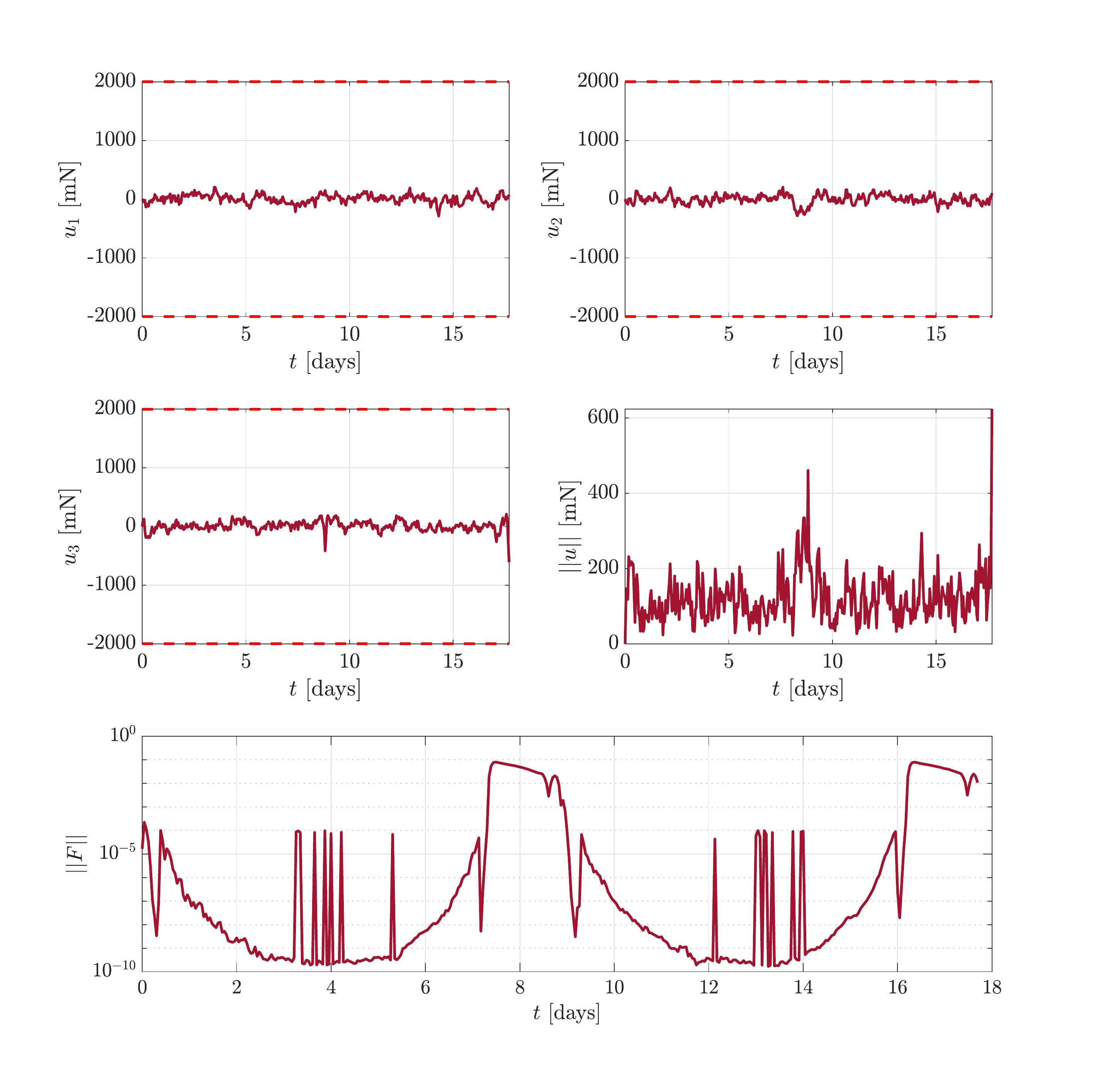}
  \caption{Control history and OCP residual $||F||$ for spacecraft in CR3BP tracking halo orbit.}
  \label{fig:control_cval}
\end{figure}

Figures \ref{fig:states_cval}--\ref{fig:control_cval} show the results of simulating the CR3BP, i.e., \eqref{eq:er3bp} with $e=0$, in closed-loop with the NMPC controller described in the Control Design Section. In this case the reference trajectory is a periodic natural motion trajectory of the CR3BP so the spacecraft is able to closely track the reference using little control input. 

\begin{figure}[htbp]
  \centering
  \includegraphics[width=0.95\textwidth]{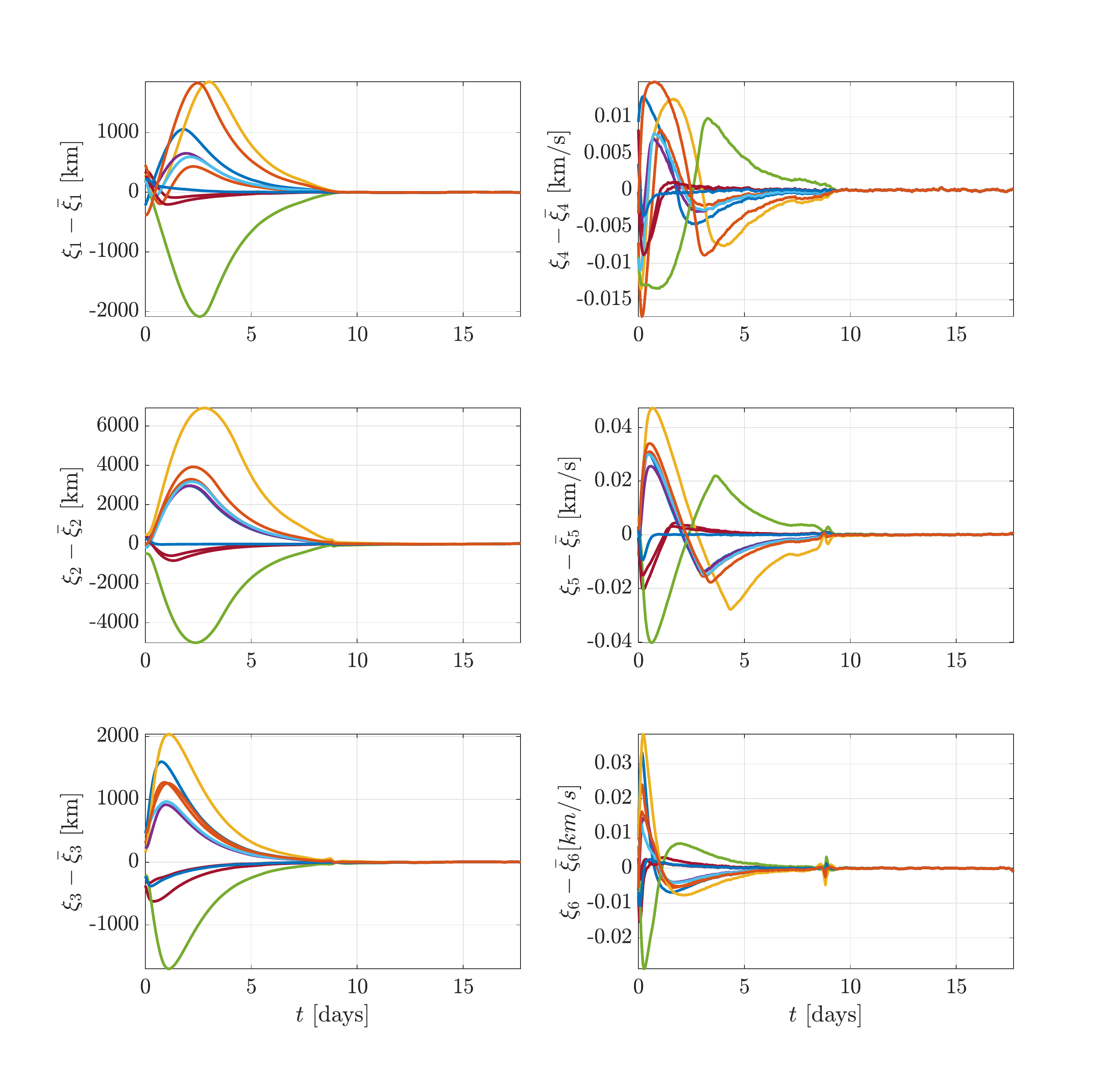}
  \caption{Spacecraft state error trajectories in the CR3BP for a variety of initial conditions.}
  \label{fig:errors_robustness}
\end{figure}

\begin{figure}[htbp]
  \centering
  \includegraphics[width=0.95\textwidth]{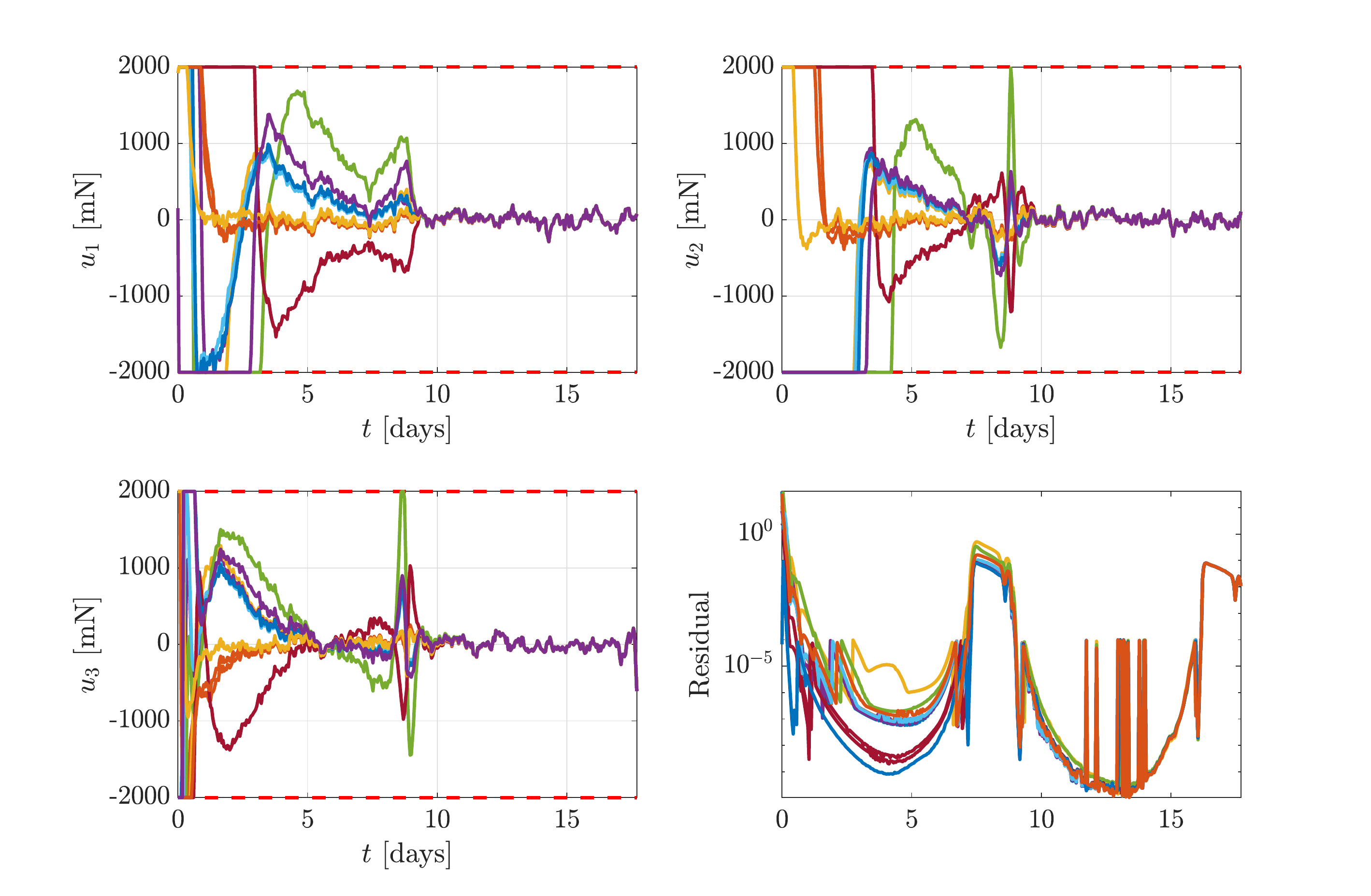}
  \caption{Spacecraft control inputs in the CR3BP for a variety of initial conditions.}
  \label{fig:control_robustness}
\end{figure}

\begin{figure}[htbp]
  \centering
  \includegraphics[width=0.95\textwidth]{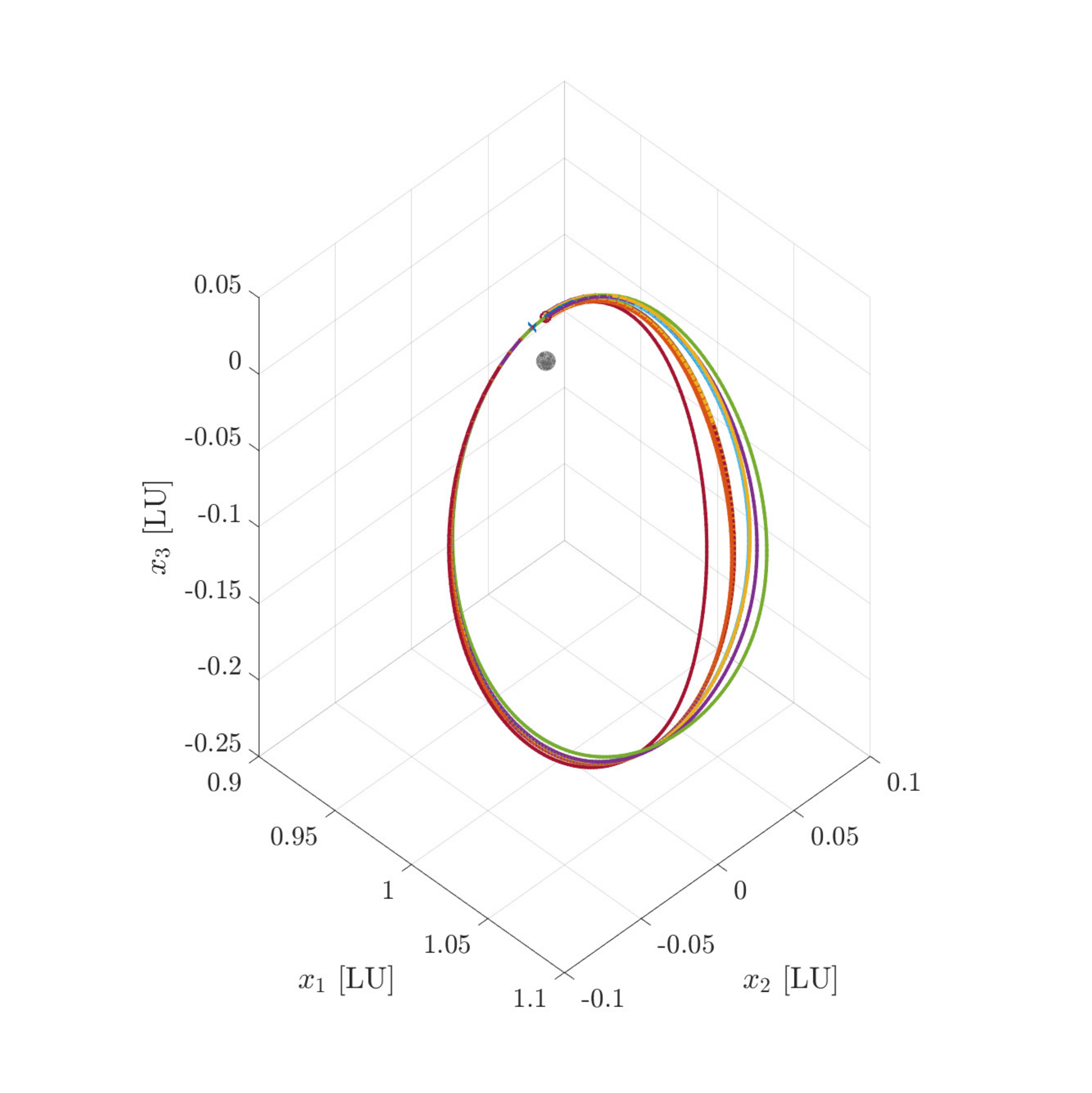}
  \caption{Spacecraft trajectory in CR3BP tracking halo orbit for various initial conditions, displayed in non-dimensional length units [LU].}
  \label{fig:robustness_3d}
\end{figure}

To test the robustness of the NMPC control strategy we performed closed-loop simulations using the CR3BP for a variety of initial conditions. To be specific, we sampled 10 initial conditions from the uniform distribution over the hypercube
\begin{equation}
  \mathcal{H} = \left \{x~|~  -\begin{bmatrix} 1_3 \Delta r_{max}\\ 1_3 \Delta v_{max} 
    
  \end{bmatrix}  \leq x-x_0 \leq \begin{bmatrix} 1_3 \Delta r_{max}\\ 1_3 \Delta v_{max} 
    
  \end{bmatrix} \right \}
\end{equation}
where $x_0 \approx (0.9878, 0, 0.0290,0 ,0.8763,0)$ is the initial condition for the Halo orbit, $\Delta r_{max} = 500~km$, and $\Delta v_{max} = 0.01~km/s$. The results are shown in Figures~\ref{fig:errors_robustness} -- \ref{fig:robustness_3d}, the spacecraft is able to reject the impact of the initial disturbance and successfully converge to the reference trajectory. As expected, this requires significant control effort.

\begin{figure}[htbp!]
	\centering
	\includegraphics[width=4in]{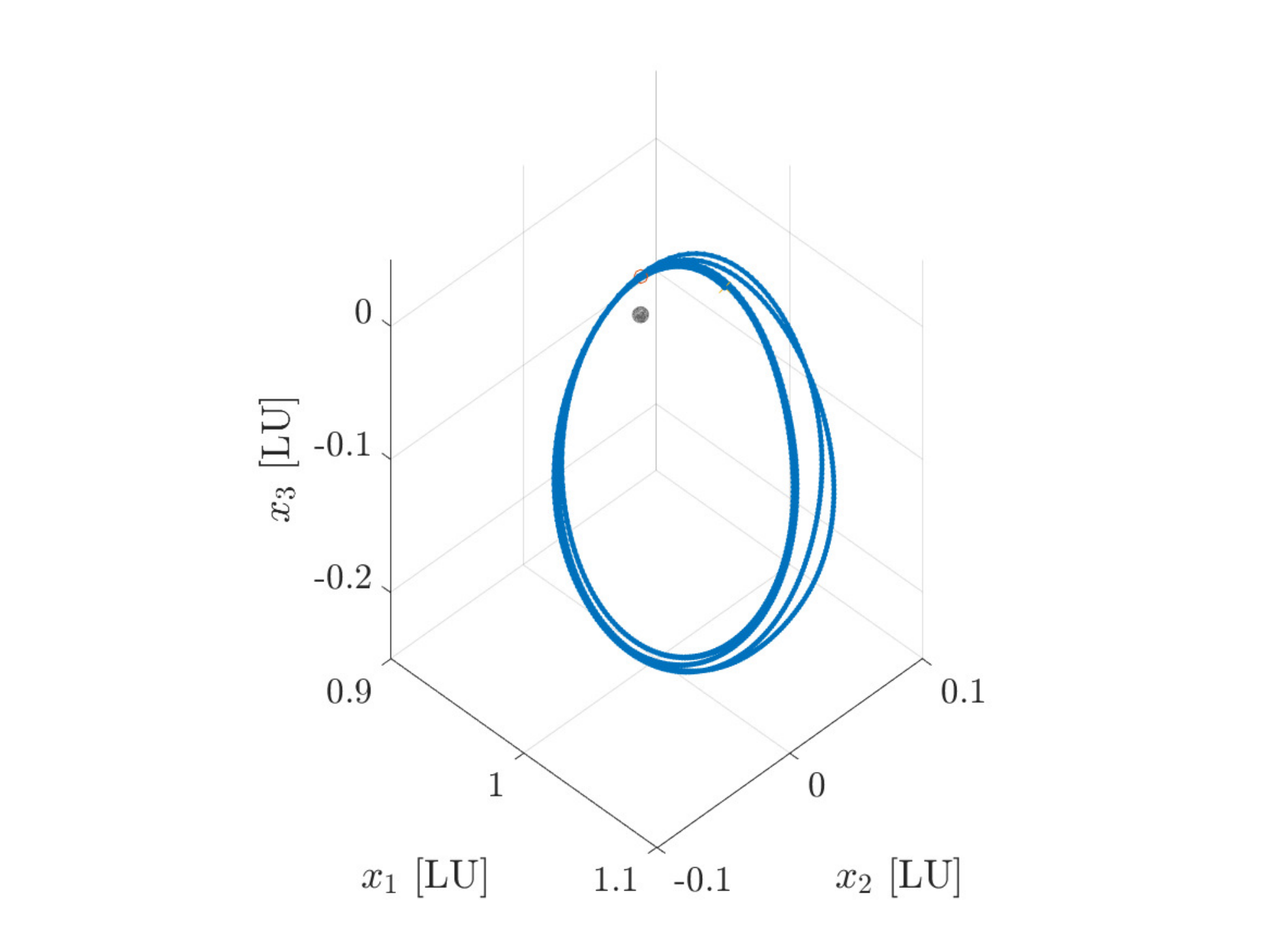}
	\caption{Spacecraft trajectory in ER3BP tracking halo orbit, displayed in non-dimensional length units [LU].}
	\label{fig:3dtraj_eval}
\end{figure}

\begin{figure}[htbp!]
	\centering
	\includegraphics[width=4in]{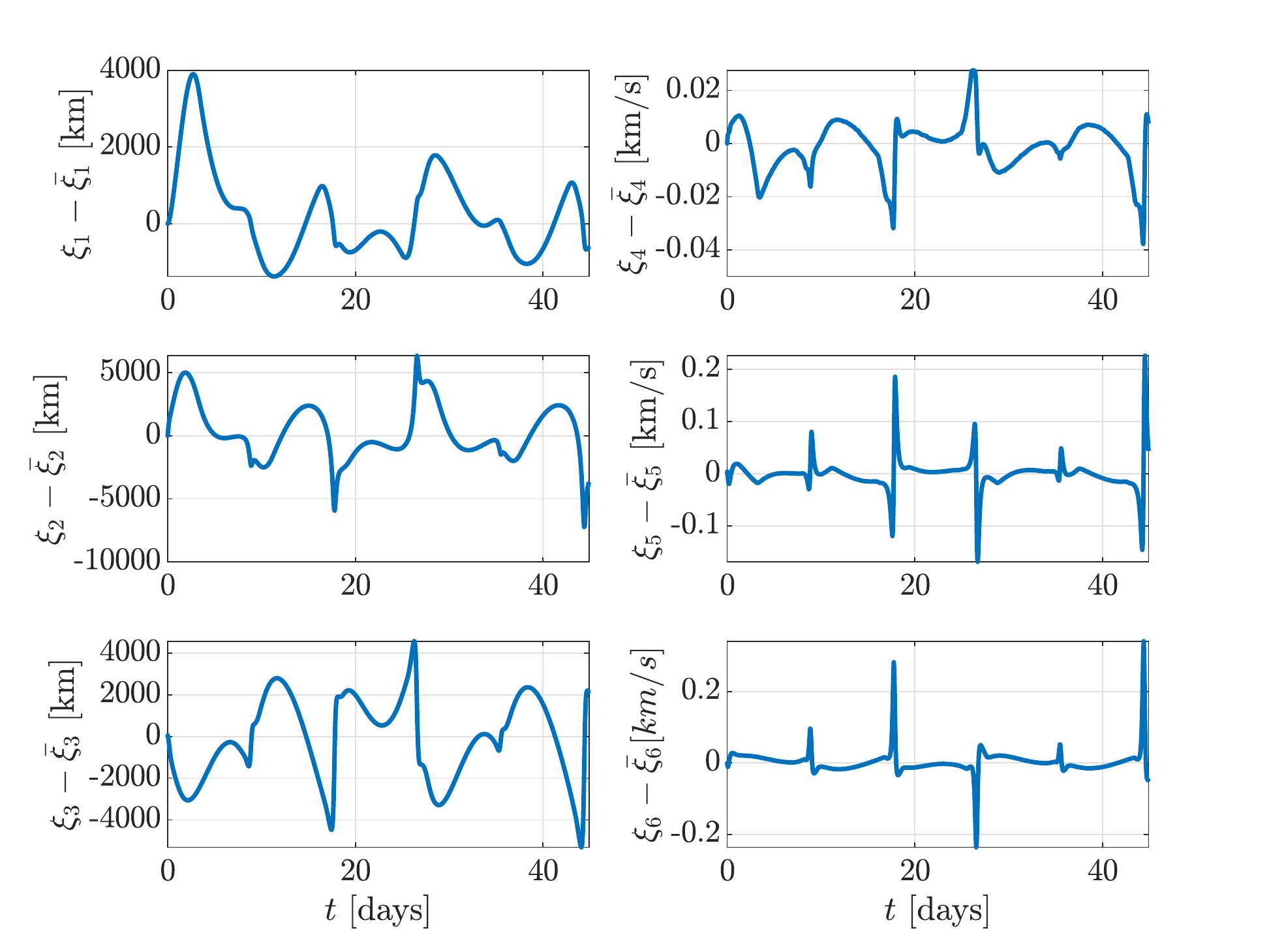}
	\caption{State error trajectories for spacecraft in ER3BP tracking halo orbit.}
	\label{fig:errors_eval}
\end{figure}

\begin{figure}[htbp!]
	\centering
	\includegraphics[width=4in]{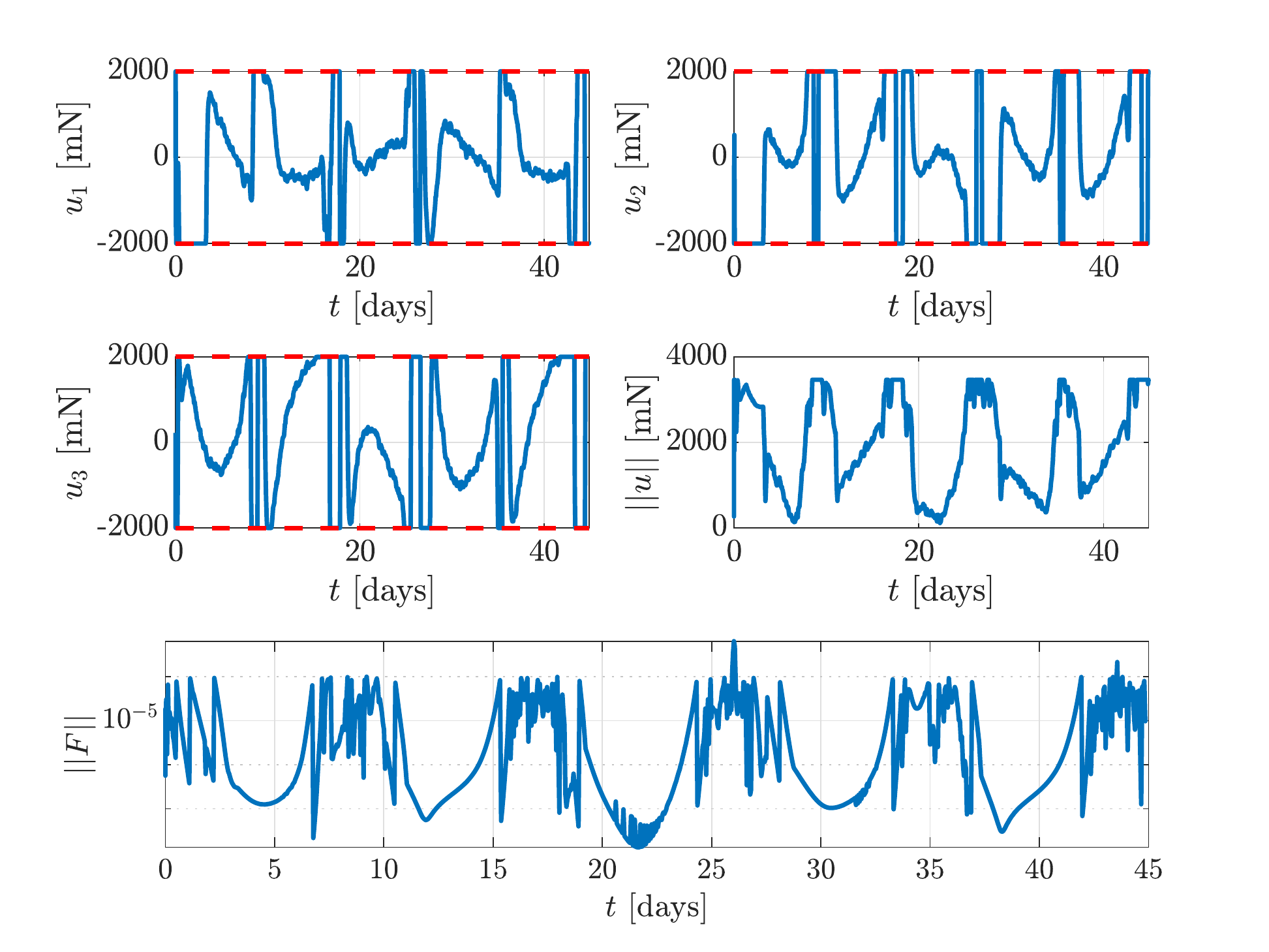}
	\caption{Control history and OCP residual $||F||$ for spacecraft in ER3BP tracking halo orbit.}
	\label{fig:control_eval}
\end{figure}

Finally, we demonstrate the robustness of the controller to model mismatch. Figures \ref{fig:3dtraj_eval} -- \ref{fig:control_eval} illustrate the closed-loop response of the controller when the simulation model eccentricity is set to $e = 0.055$. Tracking performance is significantly degraded and control effort is higher, however the controller is still able to stabilize the orbit. A comparison of Figures \ref{fig:errors_cval} and \ref{fig:errors_eval} shows the difference in tracking error between the two simulations, and a comparison of Figures \ref{fig:control_cval} and \ref{fig:control_eval} illustrates the difference in control effort and constraint handling. In the elliptic case, the spacecraft uses approximately an order of magnitude higher control effort ($2N$ vs roughly $200mN$ in the $e = 0$ case) to stabilize the system with a nonzero eccentricity. This robustness to large model mismatch and the ability to seamlessly accommodate changes in control constraints is an advantage of the MPC methodology. In the future, tracking error could be reduced by incorporating non-zero eccentricity values into the prediction model \eqref{eq:prediction_stm} and control utilization can be reduced through a better reference trajectory.

The controller has been implemented in MATLAB/Simulink using the FBstab \cite{liaomcpherson2020fbstab} quadratic programming solver and translated into \texttt{C} code by Simulink 2019a. The mean execution time of the controller found to be $8.6~ms$ on a 2019 Macbook Pro with 2.4 GHz i9 processor. A preliminary clock scaling analysis suggests a controller execution time of approximately $100~ms$ on a 200 MHz RAD750 radiation hardened computer. Given a true anomaly sampling period of $\Delta \theta = 0.01$, which implies a sampling period of approximately $1$ hour, this indicates that the computational burden of the NMPC strategy is likely to be manageable.

\begin{figure}[htbp!]
	\centering
	\includegraphics[width=4in]{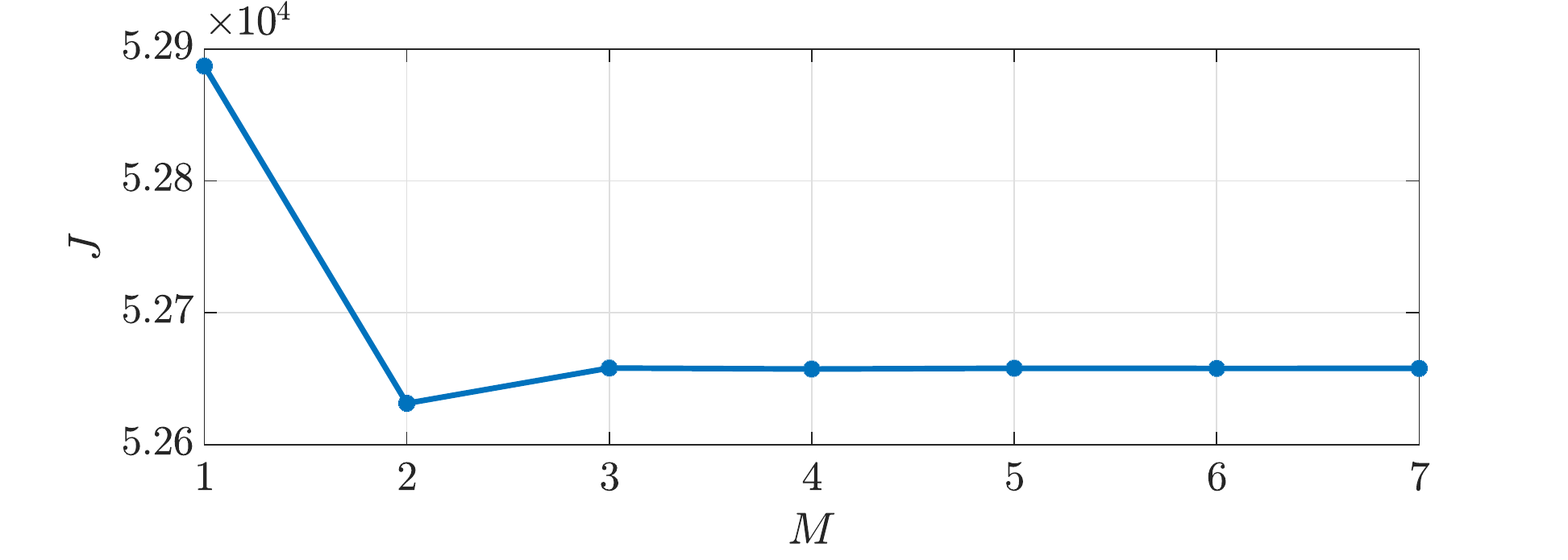}
	\caption{Sensitivity study between max SQP iterations $\ell$ and relative optimality of spacecraft trajectory tracking $J$.}
	\label{fig:TDO}
\end{figure}

In order to compare the spacecraft's performance for different values of $\ell$, we introduce a cost function $J$, analogous to \eqref{eq:OCP_cost}, 
\begin{equation} 
		J = \sum_{j} ||\xi_{j} - \bar{\xi}_{j}||_Q^2 + ||u_{j}||_R^2 
\end{equation} 
for all discrete $\theta$ instants from $\theta  = 0$ to the end of the simulation (here, 5 full revolutions of the halo orbit). The results of this sensitivity study are shown in Figure \ref{fig:TDO}, showing minimal optimality benefits to running the SQP scheme for more than three iterations. This justifies the use of $\ell=3$ and illustrates the computational efficiency benefits of the TDO scheme. 

\section{Conclusion}

In this work we have presented a nonlinear MPC-based approach to the station-keeping of halo orbits in the Earth-Moon system. A method is described for generating periodic halo orbits in the CR3BP and the same equations of motion, in combination with an RK4 integration scheme, are used in the prediction model for the controller. This nonlinear optimization problem is solved with time-distributed SQP techniques utilizing the FBstab quadratic programming algorithm. Finally, the controller is validated in numerical simulations of the CR3BP and ER3BP with process noise across several full revolutions of the halo orbit. 

The results showed that even for large disturbances and model mismatch, the NMPC scheme is capable of stabilizing the system about the reference trajectory while satisfying control constraints. Additionally, it is shown that a TDO approach to the NMPC problem allows for the use of fewer computational resources without significant reduction in performance. This work illustrates the computation feasibility of NMPC for this application which opens the door to more advanced formulations that directly incorporate high level objectives such as rendezvous or minimal propellant consumption.
\bibliographystyle{AAS_publication}   % Number the references.
\bibliography{halo_refs}   % Use references.bib to resolve the labels.

\end{document}